\documentclass[preprint]{ptephy_om}
\preprintnumber{RIKEN-iTHEMS-Report-25}

\usepackage{bm}
\usepackage{booktabs}

\numberwithin{equation}{section}

\usepackage{tikz}
\usetikzlibrary{decorations.pathmorphing}
\tikzset{snake it/.style={decorate, decoration=snake}}

\usepackage{hyperref}

\title{Unified exact WKB framework for resonance
---Zel'dovich/complex-scaling regularization and rigged~Hilbert space---}

\author[1]{Okuto Morikawa\orcid{0000-0002-0044-4491}\thanks{okuto.morikawa@riken.jp}}
\affil[1]{Center for Interdisciplinary Theoretical and Mathematical Sciences (iTHEMS),
RIKEN, Wako 351-0198, Japan}

\author[2]{Shoya Ogawa\orcid{0000-0003-0900-2486}\thanks{ogawa.shoya.615@m.kyushu-u.ac.jp}}
\affil[2]{Department of Physics, Kyushu University, 744 Motooka, Nishi-ku,
Fukuoka 819-0395, Japan}

\begin{document}
\begin{abstract}
We develop a unified framework for analyzing quantum mechanical resonances using the exact WKB method. The non-perturbative formulation based on the exact WKB method works for incorporating the Zel'dovich regularization, the complex scaling method, and the rigged Hilbert space. While previous studies have demonstrated the exact WKB analysis in bound state problems, our work extends its application to quasi-stationary states. By examining the inverted Rosen--Morse potential, we illustrate how the exact WKB analysis captures resonant phenomena in a rigorous manner. We explore the equivalence and complementarity of different well-established regularizations \`a la Zel'dovich and complex scaling within this framework. Also, we find the most essential regulator of functional analyticity and construct a modified Hilbert space of the exact WKB framework for resonance, which is called the rigged Hilbert space. This offers a deeper understanding of resonant states and their analytic structures. Our results provide a concrete demonstration of the non-perturbative accuracy of exact WKB methods in unstable quantum systems.
\end{abstract}
\maketitle

\section{Introduction}
Resonances in quantum mechanics, which correspond to unstable or quasi-stationary states, play a central role across a broad spectrum of physical contexts, including nuclear reactions, molecular scattering, optical cavities, and condensed matter excitations. Despite their ubiquity, a profound and persistent challenge has been to achieve a rigorous non-perturbative formulation of resonant states---a mathematically precise and physically transparent framework.

Recent developments in exact WKB analysis, grounded in resurgence theory, have significantly expanded the toolkit available for tackling such problems~\cite{Voros:1983xx, Delabaere:1999xx, Iwaki:2014vad}. (See Ref.~\cite{Morikawa:2025grx} and references therein.) Originally applied to bound state and stable potentials~\cite{Sueishi:2020rug,Sueishi:2021xti,Kamata:2021jrs}, this method enables non-perturbative resummation of divergent semiclassical series and offers a powerful perspective on quantum systems beyond traditional perturbative approaches. This framework has proven fruitful in the analysis of bound-state spectra and tunneling dynamics.

In this work, we aim to advance the exact WKB framework to accommodate resonant states. The domain of exact WKB theory is extended to the realm of scattering states with complex energies. Our analysis builds upon the system with the inverted Rosen--Morse potential~\cite{Zhao:1995,Landau:1991wop} in Ref.~\cite{Morikawa:2025grx}, which provided an exactly solvable yet non-trivial example of barrier resonance~\cite{Ryaboy:1993}. We embody exact WKB analysis as a unified formulation incorporating traditional and well-established approaches: the Zel'dovich regularization~\cite{Zel'dovich:1961,Berggren:1968zz} and the complex scaling method~\cite{Aguilar:1971ve,Balslev:1971vb,Myo:2014ypa}, which are historically significant and phenomenologically monumental regularization schemes. These regularizations not only allow us to compute resonances within the WKB framework but also illuminate their analytic structures through complementary approaches founded on dilation analyticity.

We demonstrate that these regularizations reveal a unifying analytic structure governing resonance physics. This leads us to identify a fundamental regulator of analyticity, inherent to any consistent regularization scheme, and motivates the construction of a rigged Hilbert space for resonance, that is, a natural extension of the conventional spectral theory~\cite{Rafael:2012,Antoniou:2001,Zhao:1995}.

Our goal is to establish the exact WKB method as a versatile and rigorous tool for resonant phenomena by elucidating the internal consistency and mutual complementarity of these methods. By clarifying the role of different regularizations and demonstrating their consistency, we open a path for future investigations not only of non-Hermitian\footnote{Recently, to non-Hermitian but $PT$-symmetric quantum mechanics, exact WKB analysis has been applied~\cite{Kamata:2023opn,Kamata:2024tyb}} and open quantum systems within a unified analytic setting, but also of the groundwork for future applications in mathematical physics, spectral theory, and quantum dynamics.

\section{Exact WKB analysis for generalized Riccati equation}
\subsection{WKB ansatz for generalized Riccati equation}
Let us first review the exact WKB analysis for a potential with first-order differentiation.\footnote{%
For more mathematical studies on first-order differentiation, see Refs.~\cite{Nikolaev:2008,Nikolaev:2021xzt}.}
A general $1$-dimensional Schr\"odinger equation is
\begin{align}
    \left[-\frac{\hbar^2}{2}\frac{d^2}{dx^2}+\frac{\hbar^2}{2} f(x)\frac{d}{dx}+V(x)\right]
    \psi(x) = E \psi(x) ,
\end{align}
where $f(x)$ is a function of~$x$ of order~$O(\hbar^0)$; in particular, under the Zel'dovich regularization, $f(x)\propto -x$.
We also have
\begin{align}
    \left[-\frac{d^2}{dx^2}+f(x)\frac{d}{dx} + \hbar^{-2}Q(x)\right]
    \psi(x) = 0 ,
\end{align}
where $Q=2(V-E)$.
Let us introduce the WKB solution of the above equation.
This is the ansatz which is defined by a formal power series as
\begin{align}
    \psi(x,\hbar) &= e^{\int^x dx'\, S(x',\hbar)} ,&
    S(x,\hbar) &= \sum_{i=-1}^{\infty}\hbar^{i} S_{i}(x) .
\end{align}
Substituting $S(x,\hbar)$ into the Schr\"odinger equation,
we have the generalized Riccati equation
\begin{align}
    S^2 + \frac{dS}{dx} - f S = \hbar^{-2} Q .
    \label{eq:riccati_eq}
\end{align}
We can obtain the recursive equation of $S_i$ for each power of~$\hbar$ as follows:
\begin{align}
    \hbar^{-2}&:S_{-1}^2 = Q ,
    \label{eq:recursive_eq_1}\\
    \hbar^{i-1}&:2S_{-1}S_{i} + \sum_{j=0}^{i-1} S_{j}S_{i-j}
    + \frac{dS_{i-1}}{dx} - f S_{i-1} = 0 \qquad i\geq0.
    \label{eq:recursive_eq}
\end{align}
From the two leading-order solutions of~Eq~\eqref{eq:recursive_eq_1}
\begin{align}
    S_{-1}(x) = S_{-1}^{\pm}(x) \equiv \pm \sqrt{Q(x)} ,
\end{align}
all higher-order terms~$S_i$ with $i\geq0$ can be determined
uniquely by Eq.~\eqref{eq:recursive_eq}.
Hence we find two solutions
\begin{align}
    S^{\pm} = \sum_{i=-1}^{\infty} \hbar^i S_i^{\pm} .
\end{align}

It is known that $S_i^{\pm}$ has the following properties:
\begin{enumerate}
    \item From Eq.~\eqref{eq:recursive_eq}, $S_i^{-}=(-1)^i S_i^{+}$ holds for $i\geq-1$ by mathematical induction;
    \item hence, immediately we can write
        \begin{align}
            S^{\pm}=\pm S_{\mathrm{odd}} + S_{\mathrm{even}},
        \end{align}
    where $S_{\mathrm{odd}}=\sum_{i\geq0}\hbar^{2i-1} S_{2i-1}^{+}$ and $S_{\mathrm{even}}=\sum_{i\geq0}\hbar^{2i} S_{2i}$;
    \item substituting the above expression into the Riccati equation~\eqref{eq:riccati_eq}, we obtain
        \begin{align}
            2 S_{\mathrm{odd}}S_{\mathrm{even}} + \left(\frac{d}{dx}-f\right) S_{\mathrm{odd}}
            = 0 .
        \end{align}
    Therefore $S_{\mathrm{even}}$ can be rewritten in terms of~$S_{\mathrm{odd}}$ by
        \begin{align}
            S_{\mathrm{even}}
            = - \frac{1}{2} \frac{1}{S_{\mathrm{odd}}}
            \left(\frac{d}{dx}-f\right) S_{\mathrm{odd}}
            = - \frac{1}{2}\frac{d}{dx} \ln S_{\mathrm{odd}}
            + \frac{1}{2}f ;
        \end{align}
    \item the WKB wave function is
        \begin{align}
            \psi(x,\hbar)^{\pm} = \frac{1}{\sqrt{S_{\mathrm{odd}}}}
            \exp\left(\pm\int^x dx'\, S_{\mathrm{odd}} + \int^x dx'\, f\right).
        \end{align}
\end{enumerate}

\subsection{Borel resummation and Stokes geometry}
The formal series, say,
\begin{align}
    \psi(x,\hbar)^{\pm}
    = e^{\pm\hbar^{-1}\int^x dx'\sqrt{Q}}
    \sum_{n\geq0} \psi_n^{\pm}(x) \hbar^{n+1/2}
\end{align}
is not necessarily convergent.
Now, we would define the Borel resummation to obtain a convergent/finite value.
The essential point is the insertion of one,
\begin{align}
    1 = \frac{1}{\Gamma(n+\alpha)} \int_0^\infty dx\, e^{-x} x^{n+\alpha-1},
\end{align}
and exchange of the infinite sum and the above integral:
\begin{align}
    \psi(x,\hbar)^{\pm}
    &= e^{\pm\hbar^{-1}\int^x dx'\sqrt{Q}}
    \sum_{n\geq0}\int_0^\infty du\, \psi_n^{\pm}(x)
    \frac{1}{\Gamma(n+1/2)}  e^{-u/\hbar} u^{n-1/2}\\
    &= \sum_{n\geq0}\int_{\mp u_0}^\infty du\, \psi_n^{\pm}(x)
    \frac{1}{\Gamma(n+1/2)}  e^{-u/\hbar} (u\pm u_0)^{n-1/2}\\
    \Psi(x,\hbar)^{\pm}
    &\equiv \int_{\mp u_0}^\infty du\, e^{-u/\hbar}
    \underbrace{\sum_{n\geq0} \frac{\psi_n^{\pm}(x)}{\Gamma(n+1/2)} (u\pm u_0)^{n-1/2}}_{=:\mathcal{B}[\psi^{\pm}](u)} ,
\end{align}
where $u_0 = \int^x dx'\sqrt{Q}$, $\psi$ is redefined as~$\Psi$,
and $\mathcal{B}[\psi^{\pm}]$ is the Borel transform.

The infinite sum in~$\mathcal{B}[\psi]$ becomes a convergent series.
On the other hand, for divergent series, the Borel transform itself may develop a singularity, called Borel singularity.
If there exists a singular point on the integration path over~$u$,
the series is Borel non-summable and the integral is ambiguous;
otherwise, this is Borel summable.
If Borel non-summable,
we may pick up some analytically continued paths
that have an imaginary difference as a non-perturbative effect.
When we vary some parameters in a system and the path jumps over the Borel singularity,
the value of the Borel integral suddenly changes.
This is the Stokes phenomenon.

It depends on the parameter~$x$ whether the WKB solution is Borel summable or not.
To see this, at first, we introduce turning points such that
\begin{align}
    \exists a,\, Q(x=a) = 0 ,
\end{align}
and a Stokes curve is defined by
\begin{align}
    \im \hbar^{-1}\int_a^x dx' \sqrt{Q} = 0. \label{eq:stokes_curve}
\end{align}
Now, we consider an integration contour from a reference point~$x_0$ to~$x$ in
\begin{align}
    u_0 = \int_{x_0}^x dx'\sqrt{Q} .
\end{align}
If this path goes across the Stokes curve,
we observe a Borel singularity with $\im u_0=0$;
it makes the integral ill-defined, i.e., Borel non-summable.
Then, the turning points and the Stokes curves determine analytically continuable regions and are called the Stokes graph or geometry.

\subsection{Connection formula near turning point}
Let us see a connection formula of the Borel resummation of the WKB solution.
Note that $\psi^{\pm}$ and $\psi^{\mp}$ are dominant and subdominant,
respectively, along a Stokes curve with
\begin{align}
    \re \hbar^{-1}\int_a^x dx' \sqrt{Q} \gtrless 0 .
\end{align}
Let regions I and II be Borel summable areas,
whose boundary curve starting from the same turning point makes $\psi^{\pm}$ dominant.
When we go across this Stokes curve from I to II near the turning point, the Stokes phenomenon gives rise to
\begin{align}
    \begin{pmatrix}
      \psi^{+}_{\mathrm{I}} \\ \psi^{-}_{\mathrm{I}}
    \end{pmatrix}
    =
    M
    \begin{pmatrix}
      \psi^{+}_{\mathrm{II}} \\ \psi^{-}_{\mathrm{II}}
    \end{pmatrix} ,
    \label{eq:connection}
\end{align}
where $M$ depends on the rotation of the path around the turning point,
\begin{align}
    M =
    \begin{cases}
        M_{+} & \text{if anti-clockwise for the index~$+$,} \\
        M_{+}^{-1}& \text{if clockwise for the index~$+$,} \\
        M_{-} & \text{if anti-clockwise for the index~$-$,} \\
        M_{-}^{-1} &\text{if clockwise for the index~$-$,}
    \end{cases}
\end{align}
and
\begin{align}
    M_{+} &=
    \begin{pmatrix}
        1 & i \\ 0 & 1
    \end{pmatrix}, &
    M_{-} &=
    \begin{pmatrix}
        1 & 0 \\ i & 1
    \end{pmatrix} .
\end{align}
We can connect the different turning points, $a_1$ and $a_2$,
such that the wave function is related to each other by the factor,
\begin{align}
    N_{a_1a_2}=\diag\left(\exp\left(\int_{a_1}^{a_2}dx\, S_{\mathrm{odd}} + \int_{a_1}^{a_2} dx\, f\right),\exp\left(-\int_{a_1}^{a_2}dx\, S_{\mathrm{odd}} + \int_{a_1}^{a_2} dx\, f\right)\right) .
\end{align}
Noting that there exists a branch cut due to $1/\sqrt{S_{\mathrm{odd}}}$, after crossing this cut, i.e., going to another Riemann sheet, the index $+$ ($-$) must be replaced by $-$ ($+$) (see Fig.~\ref{fig:branch}).

\begin{figure}
\centering
  \begin{tikzpicture}[scale=0.12]
   \draw[line width=1pt] (22,25) .. controls (22,0) and (30,0) .. (30,-25);
   \draw[line width=1pt] (30,28) .. controls (30,0) and (22,0) .. (22,-28);
   \draw[line width=1pt] (22,25) -- (-30,25) -- (-30,-28) -- (22,-28);
   \draw[line width=1pt] (30,28) -- (-22,28) -- (-22,25);
   \draw[line width=1pt] (30,-25) -- (22,-25);
   \draw[line width=2pt] (-4,0) .. controls (-10,-3) and (-22,-3) .. (-27,-3) node[below] {$+$};
   \draw[gray,dashed,line width=2pt] (-4,0) .. controls (-10,3) and (-22,3) .. (-27,3) node[above] {$-$};
   \draw[line width=2pt] (-4,0) .. controls (-7,-10) and (-7,-22) .. (-7,-24-3) node[left] {$-$};
   \draw[gray,dashed,line width=2pt] (-4,0) .. controls (-1,-10) and (-1,-22) .. (-1,-24-3) node[right] {$+$};
   \draw[line width=2pt] (-4,0) .. controls (-7,10) and (-7,22) .. (-7,24) node[left] {$-$};
   \draw[gray,dashed,line width=2pt] (-4,0) .. controls (-1,10) and (-1,22) .. (-1,24) node[right] {$+$};
   \draw[blue,line width=3pt,snake it] (-4,0) -- (26,0);
   \draw[line width=1pt] (-26,17) -- (-21,17) -- (-21,22);
   \draw[line width=1pt] (-23.5,19.5) node {$x$};
  \end{tikzpicture}
  \caption{Stokes graph near a turning point. The solid and dashed curves are the associated Stokes curves on the first and second Riemann sheets, respectively. The blue wavy line is a branch cut. After crossing a branch cut once, the correspondence to the indices $+$ and $-$ in $M$ must be reversed.}
  \label{fig:branch}
\end{figure}

\section{Zel'dovich regularization}
\subsection{Hamiltonian under Zel'dovich transformation}
The wave function of a resonant state diverges in the asymptotic region.
This singular nature causes difficulty in investigating the state as follows: In the $3$-dimensional space, the radial wave function $\varphi(r)$ of a resonant state with $k=k_{r}-ik_{i}$ ($k_{r}$, $k_{i}\in\mathbb{R}^{+}$) has the typical asymptotic form
\begin{align}
 \varphi(r) \xrightarrow[r\to\infty]{}e^{ikr} = e^{ik_{r}}e^{k_{i}r},
\end{align}
which grows up exponentially because of~$e^{k_{i}r}$.
The integration of the norm, $||\varphi(r)||^2$, is also ill-defined.
Similarly, in time-independent scattering theory,
a factor, $e^{i\varepsilon t}$ with $\varepsilon>0$, suppress the $e^{-iEt}$ oscillations in stationary solutions.
An analogous approach, called the Zel'dovich regularization~\cite{Zel'dovich:1961}, can be introduced in order to obtain a finite value of the norm as
\begin{align}
 ||\varphi(r)||_S^2\equiv\lim_{\varepsilon\to+0}\int_{0}^{\infty} dr\, e^{-\varepsilon r^2}\varphi(r)^2r^2 .
\end{align}
Using this regularization, Berggren~\cite{Berggren:1968zz} proved the orthogonality and completeness properties of the resonant states.
Now, the square of the wave function is integrated with the norm not only of the bound states but also of the resonant states.

For the Schr\"odinger equation,
we can have a similarity transformation associated with Zel'dovich regularization.
The Zel'dovich transformation is given by
\begin{align}
    H\psi = E\psi
    \qquad\Rightarrow\qquad
    H_S\psi_S = E\psi_S
\end{align}
with
\begin{align}
    \psi_S &\equiv S\psi,& H_S &\equiv S H S^{-1}.
\end{align}
Here $S$ is the Gaussian factor $e^{-\varepsilon r^2}$.
Note that the Zel'dovich-transformed Hamiltonian~\cite{Moiseyev:2011} is related to the original one as
\begin{align}
    H_S = H - \frac{\hbar^2\varepsilon}{m}(\nabla \cdot \bm{r} + \bm{r} \cdot \nabla)
    + O(\varepsilon^2) .
\end{align}
For the $1$-dimensional system, one finds that
\begin{align}
    H_S = H - \frac{\hbar^2\varepsilon}{m} \left(\frac{d}{dx}x+x\frac{d}{dx}\right)
        = H - \frac{\hbar^2\varepsilon}{m} \left(1 + 2 x \frac{d}{dx}\right).
    \label{eq:zel_hamiltonian}
\end{align}

\subsection{Inverted Rosen--Morse potential}
The Stokes graph via the Zel'dovich transformation is identical to the original graph
with $f(x)=-\frac{4\varepsilon}{m}x$ in Eq.~\eqref{eq:zel_hamiltonian}.
For the inverted Rosen--Morse potential,
\begin{align}
    V(x) = \frac{U_0}{\cosh^2\beta x},
        \qquad U_0 > 0,
\end{align}
see Figs.~\ref{fig:stokes_graph} ($\im\hbar\gtrless0$ and~$\im E\gtrless0$) and \ref{fig:stokes_graph2} ($\im\hbar\gtrless0$ and~$\im E\lessgtr0$, but $\im E\sim0$) \cite{Morikawa:2025grx}.
Note that $\im E_{\mathrm{res}}<0$ for the complex energy of resonance, $E_{\mathrm{res}}$.
Now, since the wave function is expected to be normalizable after the Zel'dovich transformation,
Note that the normalizability of the wave function,
i.e., the dominance of~$\psi^{-}$ at infinity,
gives rise to the quantization condition which is given by the analytical continuation and the connection formula.
One may choose the simple path depicted on the left panel in~Fig.~\ref{fig:stokes_graph_path}
for giving the quantization condition.
The non-perturbative cycle, $A=e^{\oint_{a_1}^{a_2}dx\,S_{\text{odd}}}$, contributes $M_{-}N_{a_1a_2}M_{+}$, where $a_{1,2}$ is a turning point; a simple convergence condition that $\psi^{+}$ should vanish at~$\im x\to\pm\infty$ gives rise to $1-A=0$ as is given in Ref.~\cite{Morikawa:2025grx}.
\begin{figure}[t]
\centering
\begin{tikzpicture}[scale=1.2]
  \draw[->] (-2.5,-1.4) -- (2.5,-1.4) node[right] {$\re x$};
  \draw[->] (0,-3) -- (0,3) node[above] {$\im x$};
  \draw[thick,domain=0:520,samples=100,variable=\t,rotate=20] plot ({-0.002*\t*sin(\t)}, {-0.0018*\t^1.05*cos(\t)}) ..controls (0.5,1.35) and (1,1.1).. (1.4,1) node[above left] {$-$};
  \draw[thick,domain=0:-520,samples=100,variable=\t,rotate=20] plot ({0.002*\t*sin(\t)}, {-0.0018*\t^1.05*cos(\t)}) ..controls (-0.5,-1.35) and (-1,-1.1).. (-1.4,-1) node[below right] {$+$};
  \draw[thick,domain=0:550,samples=100,variable=\t,rotate=110] plot ({-0.0018*\t^1.05*sin(\t)}, {-0.002*\t*cos(\t)}) ..controls (1.5,0.9) and (1.5,0.4).. (1.7,0.5) node[below left] {$+$};
  \draw[thick,domain=0:-550,samples=100,variable=\t,rotate=110] plot ({0.0018*\t^1.05*sin(\t)}, {-0.002*\t*cos(\t)}) ..controls (-1.5,-0.9) and (-1.5,-0.4).. (-1.7,-0.5) node[above right] {$-$};
  \draw[dashed] (1,1.4) ..controls (1.15,1.5) and (1.2,2) .. (0.9,2.8);
  \draw[dashed] (-1,-1.4) ..controls (-1.15,-1.5) and (-1.2,-2) .. (-0.9,-2.8);
  \draw[dashed] (-1,1.4)   ..controls (-0.5,1.4)   and (-0.2,1.6)  .. (0,1.7);
  \draw[dashed] (1,-1.4) ..controls (0.5,-1.4) and (0.2,-1.6) .. (0,-1.7);
  \draw[thick] (1,1.4) ..controls (1.5,1) and (1.6,-1.4) .. (1.5,-2.8) node[right] {$+$};
  \draw[thick] (-1,-1.4) ..controls (-1.5,-1) and (-1.6,1.4) .. (-1.5,2.8) node[left] {$-$};
  \fill[blue] (0,0) circle(4pt);
  \draw[very thick,blue, snake it] (-1,1.4) -- (1,-1.4);
  \fill (-1,1.4) circle(3pt); \fill (1,1.4) circle(3pt);
  \fill (1,-1.4) circle(3pt); \fill (-1,-1.4) circle(3pt);
\end{tikzpicture}
\hspace{1em}
\begin{tikzpicture}[scale=1.2]
  \draw[->] (-2.5,-1.4) -- (2.5,-1.4) node[right] {$\re x$};
  \draw[->] (0,-3) -- (0,3) node[above] {$\im x$};
  \fill (-1,1.4) circle(3pt); \fill (1,1.4) circle(3pt);
  \fill (1,-1.4) circle(3pt); \fill (-1,-1.4) circle(3pt);
  \draw[thick,domain=0:-520,samples=100,variable=\t,rotate=-20] plot ({0.002*\t*sin(\t)}, {0.0018*\t^1.05*cos(\t)}) ..controls (-0.5,1.35) and (-1,1.1).. (-1.4,1) node[above right] {$+$};
  \draw[thick,domain=0:520,samples=100,variable=\t,rotate=-20] plot ({-0.002*\t*sin(\t)}, {0.0018*\t^1.05*cos(\t)}) ..controls (0.5,-1.35) and (1,-1.1).. (1.4,-1) node[below left] {$-$};
  \draw[thick,domain=0:-550,samples=100,variable=\t,rotate=-110] plot ({0.0018*\t^1.05*sin(\t)}, {0.002*\t*cos(\t)}) ..controls (-1.5,0.9) and (-1.5,0.4).. (-1.7,0.5) node[below right] {$-$};
  \draw[thick,domain=0:550,samples=100,variable=\t,rotate=-110] plot ({-0.0018*\t^1.05*sin(\t)}, {0.002*\t*cos(\t)}) ..controls (1.5,-0.9) and (1.5,-0.4).. (1.7,-0.5) node[above left] {$+$};
  \draw[dashed] (-1,1.4) ..controls (-1.15,1.5) and (-1.2,2) .. (-0.9,2.8);
  \draw[dashed] (1,-1.4) ..controls (1.15,-1.5) and (1.2,-2) .. (0.9,-2.8);
  \draw[dashed] (1,1.4)   ..controls (0.5,1.4)   and (0.2,1.6)   .. (0,1.7);
  \draw[dashed] (-1,-1.4) ..controls (-0.5,-1.4) and (-0.2,-1.6) .. (0,-1.7);
  \draw[thick] (-1,1.4) ..controls (-1.5,1) and (-1.6,-1.4) .. (-1.5,-2.8) node[left] {$-$};
  \draw[thick] (1,-1.4) ..controls (1.5,-1) and (1.6,1.4) .. (1.5,2.8) node[right] {$+$};
  \fill[blue] (0,0) circle(4pt);
  \draw[very thick,blue, snake it] (1,1.4) -- (-1,-1.4);
\end{tikzpicture}
\caption{Schematic illustration of Stokes graph for the inverted Rosen--Morse potential, $V(x)=1/\cosh^2x$, via the Zel'dovich transformation.
This geometry itself is identical to the original one~\cite{Morikawa:2025grx}.
The left/right panels are devoted to~$\im\hbar\gtrless0$ and~$\im E\gtrless0$.
The black points are the turning points and the solid curves are the associated Stokes curves.
The dashed Stokes curves mean the periodicity of~$\im x\in[-\pi/2,\pi/2]$ due to the cosine function on~$\re x=0$.
The blue points are double poles and the blue wavy lines are branch cuts.}
\label{fig:stokes_graph}
\end{figure}
\begin{figure}[t]
\centering
\begin{tikzpicture}[scale=1.2]
  \draw[->] (-2.5,-1.4) -- (2.5,-1.4) node[right] {$\re x$};
  \draw[->] (0,-3) -- (0,3) node[above] {$\im x$};
  %
  \draw[thick,domain=0:550,samples=100,variable=\t,rotate=110] plot ({-0.0018*\t^1.05*sin(\t)}, {-0.002*\t*cos(\t)}) ..controls (1.5,0.9) and (1.5,0.4).. (1.7,0.5) node[below left] {$+$};
  \draw[thick,domain=0:-550,samples=100,variable=\t,rotate=110] plot ({0.0018*\t^1.05*sin(\t)}, {-0.002*\t*cos(\t)}) ..controls (-1.5,-0.9) and (-1.5,-0.4).. (-1.7,-0.5) node[above right] {$-$};
  \draw[very thick] (-1,-1.4) -- (1,-1.4);
  \draw[very thick] (-1,1.4) -- (1,1.4);
  \draw[dashed] (1,1.4) ..controls (1.15,1.5) and (1.2,2) .. (0.9,2.8);
  \draw[dashed] (-1,-1.4) ..controls (-1.15,-1.5) and (-1.2,-2) .. (-0.9,-2.8);
  %
  \draw[thick] (1,1.4) ..controls (1.5,1) and (1.6,-1.4) .. (1.5,-2.8) node[right] {$+$};
  \draw[thick] (-1,-1.4) ..controls (-1.5,-1) and (-1.6,1.4) .. (-1.5,2.8) node[left] {$-$};
  \fill[blue] (0,0) circle(4pt);
  \draw[very thick,blue, snake it] (-1,1.4) -- (1,-1.4);
  \fill (-1,1.4) circle(3pt); \fill (1,1.4) circle(3pt);
  \fill (1,-1.4) circle(3pt); \fill (-1,-1.4) circle(3pt);
\end{tikzpicture}
\hspace{1em}
\begin{tikzpicture}[scale=1.2]
  \draw[->] (-2.5,-1.4) -- (2.5,-1.4) node[right] {$\re x$};
  \draw[->] (0,-3) -- (0,3) node[above] {$\im x$};
  \fill (-1,1.4) circle(3pt); \fill (1,1.4) circle(3pt);
  \fill (1,-1.4) circle(3pt); \fill (-1,-1.4) circle(3pt);
  %
  \draw[thick,domain=0:-550,samples=100,variable=\t,rotate=-110] plot ({0.0018*\t^1.05*sin(\t)}, {0.002*\t*cos(\t)}) ..controls (-1.5,0.9) and (-1.5,0.4).. (-1.7,0.5) node[below right] {$-$};
  \draw[thick,domain=0:550,samples=100,variable=\t,rotate=-110] plot ({-0.0018*\t^1.05*sin(\t)}, {0.002*\t*cos(\t)}) ..controls (1.5,-0.9) and (1.5,-0.4).. (1.7,-0.5) node[above left] {$+$};
  \draw[very thick] (-1,-1.4) -- (1,-1.4);
  \draw[very thick] (-1,1.4) -- (1,1.4);
  \draw[dashed] (-1,1.4) ..controls (-1.15,1.5) and (-1.2,2) .. (-0.9,2.8);
  \draw[dashed] (1,-1.4) ..controls (1.15,-1.5) and (1.2,-2) .. (0.9,-2.8);
  %
  \draw[thick] (-1,1.4) ..controls (-1.5,1) and (-1.6,-1.4) .. (-1.5,-2.8) node[left] {$-$};
  \draw[thick] (1,-1.4) ..controls (1.5,-1) and (1.6,1.4) .. (1.5,2.8) node[right] {$+$};
  \fill[blue] (0,0) circle(4pt);
  \draw[very thick,blue, snake it] (1,1.4) -- (-1,-1.4);
\end{tikzpicture}
\caption{Schematic illustration of Stokes graph for the inverted Rosen--Morse potential. Same as Fig.~\ref{fig:stokes_graph}, but $\im\hbar\gtrless0$ and~$\im E\lessgtr0$. For simplicity, $\im E\sim 0$.}
\label{fig:stokes_graph2}
\end{figure}

\begin{figure}[t]
\centering
\begin{tikzpicture}[scale=1.4]
  \draw[->] (-2.5,-1.4) -- (2,-1.4) node[right] {$\re x$};
  \draw[->] (0,-3.2) -- (0,1.1) node[above] {$\im x$};
  \draw (0,-8) -- (0,-4); \draw[dashed] (0,-4) -- (0,-3.2);
  \draw[thick] (1,-1.4) ..controls (1.15,-1.3) and (1.2,-0.8) .. (0.9,0) node[above] {$-$};
  \draw[thick] (-1,-1.4) ..controls (-1.15,-1.5) and (-1.2,-2) .. (-0.9,-2.8) node[left] {$+$};
  \draw[very thick] (-1,-1.4) -- (1,-1.4);
  \draw[thick] (1.2,0.5) ..controls (1.4,0) and (1.6,-1.4) .. (1.5,-2.8) node[right] {$+$};
  \draw[thick] (-1,-1.4) ..controls (-1.5,-1) and (-1.6,1) .. (-1.6,1) node[left] {$-$};
  \draw[thick] (1,-1.4) ..controls (1.4,-1.6) and (1.3,-2.5) .. (1.3,-2.8) node[left] {$+$};
  \fill[blue] (0,0) circle(3pt); \fill[blue] (0,-2.8) circle(3pt);
  \draw[blue, snake it] (0,0) -- (1,-1.4);
  \draw[blue, snake it] (0,-2.8) -- (-1,-1.4);
  \fill (1,-1.4) circle(3pt); \fill (-1,-1.4) circle(3pt);
  \draw[very thick,dashed,teal] (0,-1.4) circle [x radius=1.3,y radius=0.4] node[below left] {$A$};
  %
  \draw[thick] (1,-6.4) ..controls (1.15,-6.3) and (1.2,-5.8) .. (0.9,-5) node[above] {$-$};
  \draw[thick] (-1,-6.4) ..controls (-1.15,-6.5) and (-1.2,-7) .. (-0.9,-7.8) node[left] {$+$};
  \draw[very thick] (-1,-6.4) -- (1,-6.4);
  \draw[thick] (1.2,-4.5) ..controls (1.4,-5) and (1.6,-6.4) .. (1.5,-7.8) node[right] {$+$};
  \draw[thick] (-1,-6.4) ..controls (-1.5,-6) and (-1.6,-4) .. (-1.6,-4) node[left] {$-$};
  \draw[thick] (1,-6.4) ..controls (1.4,-6.6) and (1.3,-7.5) .. (1.3,-7.8) node[left] {$+$};
  \fill[blue] (0,-5) circle(3pt); \fill[blue] (0,-7.8) circle(3pt);
  \draw[blue, snake it] (0,-5) -- (1,-6.4);
  \draw[blue, snake it] (0,-7.8) -- (-1,-6.4);
  \fill (1,-6.4) circle(3pt); \fill (-1,-6.4) circle(3pt);
  %
  \draw[ultra thick,red] (-2.5,-1) -- (0.65,-1);
  \draw[ultra thick,dashed,red,->] (0.6,-1) -- (2.5,-1);
\end{tikzpicture}
\hspace{1em}
\begin{tikzpicture}[scale=1.4]
  \draw[->] (-2.5,-1.4) -- (2,-1.4) node[right] {$\re x$};
  \draw[->] (0,-3.2) -- (0,1.1) node[above] {$\im x$};
  \draw (0,-8) -- (0,-4); \draw[dashed] (0,-4) -- (0,-3.2);
  \draw[thick] (1,-1.4) ..controls (1.15,-1.3) and (1.2,-0.8) .. (0.9,0) node[above] {$-$};
  \draw[thick] (-1,-1.4) ..controls (-1.15,-1.5) and (-1.2,-2) .. (-0.9,-2.8) node[left] {$+$};
  \draw[very thick] (-1,-1.4) -- (1,-1.4);
  \draw[thick] (1.2,0.5) ..controls (1.4,0) and (1.6,-1.4) .. (1.5,-2.8) node[right] {$+$};
  \draw[thick] (-1,-1.4) ..controls (-1.5,-1) and (-1.6,1) .. (-1.6,1) node[left] {$-$};
  \draw[thick] (1,-1.4) ..controls (1.4,-1.6) and (1.3,-2.5) .. (1.3,-2.8) node[left] {$+$};
  \fill[blue] (0,0) circle(3pt); \fill[blue] (0,-2.8) circle(3pt);
  \draw[blue, snake it] (0,0) -- (1,-1.4);
  \draw[blue, snake it] (0,-2.8) -- (-1,-1.4);
  \fill (1,-1.4) circle(3pt); \fill (-1,-1.4) circle(3pt);
  \draw[very thick,dashed,teal] (0,-1.4) circle [x radius=1.3,y radius=0.4] node[below left] {$A$};
  %
  \draw[thick] (1,-6.4) ..controls (1.15,-6.3) and (1.2,-5.8) .. (0.9,-5) node[above] {$-$};
  \draw[thick] (-1,-6.4) ..controls (-1.15,-6.5) and (-1.2,-7) .. (-0.9,-7.8) node[left] {$+$};
  \draw[very thick] (-1,-6.4) -- (1,-6.4);
  \draw[thick] (1.2,-4.5) ..controls (1.4,-5) and (1.6,-6.4) .. (1.5,-7.8) node[right] {$+$};
  \draw[thick] (-1,-6.4) ..controls (-1.5,-6) and (-1.6,-4) .. (-1.6,-4) node[left] {$-$};
  \draw[thick] (1,-6.4) ..controls (1.4,-6.6) and (1.3,-7.5) .. (1.3,-7.8) node[left] {$+$};
  \fill[blue] (0,-5) circle(3pt); \fill[blue] (0,-7.8) circle(3pt);
  \draw[blue, snake it] (0,-5) -- (1,-6.4);
  \draw[blue, snake it] (0,-7.8) -- (-1,-6.4);
  \fill (1,-6.4) circle(3pt); \fill (-1,-6.4) circle(3pt);
  %
  \draw[ultra thick,red] (-1.3,-8) -- (-1.3,-1) -- (0.6,-1);
  \draw[ultra thick,red,dashed] (0.6,-1) -- (1.3,-1) -- (1.3,1);
  \draw[ultra thick,dotted,red] (-2.6,-1) .. controls (-2.6,-5) and (-1.8,-7) .. (-1.3,-8); \node[red,above] at (-2.6,-1) {$-\infty$};
  \draw[ultra thick,dotted,red,->] (1.3,1) .. controls (1.8,0.5) and (2.5,0) .. (2.5,-0.8) node[below] {$+\infty$};
\end{tikzpicture}
\caption{Quantization condition with~$\im\hbar>0$ and $\im E\lesssim0$ for resonance.
The red path on the left panel is a simple option to carry out the analytical continuation,
along which we can expect the normalizability of the exact WKB solution.
The right panel shows actual connection manipulations of it.
The red solid line is on the Riemann sheet depicted and the red dashed line is on another Riemann sheet after passing through the branch cut;
the red dotted curve is the analytical continuation at infinity.
The nontrivial cycle, $A$-cycle, is shown as the green loop.}
\label{fig:stokes_graph_path}
\end{figure}

We should now observe the convergence of (norm of) the wave function at~$x=\pm\infty$.
Then, the actual connection path is given by the red path on the right panel in Fig.~\ref{fig:stokes_graph_path}.
The red solid line is on the Riemann sheet depicted and the red dashed line is on another Riemann sheet after passing through the branch cut.
In the WKB solution,
$e^{-\int^x dx'\, S_{\mathrm{odd}}}$ along the Stokes curve with the index~$-$ exponentially dumps,
and hence the dominant effect comes from~$e^{\int^x dx'\, f}=e^{-2\varepsilon x^2/m}$.
At $\im x=\pm\infty$, the overall factor of the WKB solution diverges exponentially,
while at $\re x=\pm\infty$ it converges exponentially.
As a consequence, we find that this consistency provides the connection formula for $\psi=(\psi^{+},\psi^{-})^T$ between $x\to\pm\infty$ (the following analytic continuation at infinity is carried out in each Borel summable region)
\begin{align}
    \psi|_{x\to-\infty} &= \left[\text{analytic continuation from $-\infty$ to $\lim_{n\to\infty}(-a-2\pi i n/\beta)$}\right]\notag\\
    &\qquad
    \times \prod_{n=\infty}^{0} M_{-}^{-1} N_{-a-2\pi i (n+1)/\beta,-a-2\pi i n/\beta}\notag\\
    &\qquad
    \times M_{-}^{-1} N_{-a,a} M_{+}^{-1} \notag\\
    &\qquad
    \times \prod_{n=0}^{\infty} N_{a+2\pi i n/\beta,a+2\pi i (n+1)/\beta} M_{-}\notag\\
    &\qquad
    \times \left[\text{analytic continuation from $\lim_{n\to\infty}(a+2\pi i n/\beta)$ to $\infty$}\right] \psi|_{x\to\infty} .
\end{align}
Hence, $\psi^{\pm}|_{x\to-\infty}$ is a linear combination of $\psi^+|_{x\to\infty}$ and $\psi^-|_{x\to\infty}$.
Reference~\cite{Morikawa:2025grx} states that the resonant wave-function is described by the connection formula of $\psi^{-}$ from~$\im x\to\infty$ to $\im x\to-\infty$. From the observation of the explicit asymptotic behavior of the exact solution (see Section~4.2.2 in Ref.~\cite{Morikawa:2025grx}), the analytical-continued wave-function $\psi^-|_{x\to\infty}$ corresponds to the wave function being normalizable on a path extended in the $\im x$-direction, but is divergent on the current horizontal path in the $\re x$-direction as the non-normalizability problem of resonance. Then, the analytically continued $\psi^+|_{x\to-\infty}$ is an indeed non-normalizable solution.\footnote{Note that the current path in this paper is similar near $x=a$ to that in Ref.~\cite{Morikawa:2025grx}, but is quite different near $x=-a$ from it. Then the latter path reverses the asymptotic behavior of~$\psi^{\pm}$.} Now, noting that
\begin{align}
    \cdots M_{-}^{-1} N_{-a,a} M_{+}^{-1} \cdots \propto
    \begin{pmatrix}
        1 & - i \\ - i & \frac{1}{A} - 1
    \end{pmatrix},
\end{align}
where $A$ is the non-perturbative cycle as~$A=e^{\oint_{-a}^{a}dx\, S_{\mathrm{odd}}}$ and the dots ``$\cdots$'' do not alter the structure of the matrix,
the coefficient of $\psi^{-}$, the $(2,2)$ entry, must vanish due to the consistency of blowing-up behaviors of $\psi^-|_{x\to\infty}$ and $\psi^+|_{x\to-\infty}$.
To satisfy this normalizability of the wave function,
the quantization condition is given by
\begin{align}
    0 &=e^{\int_{a+i\infty}^{a+\infty} dx (-S_{\mathrm{odd}}+f)}
    \left[\prod_{n=0}^{\infty}
    e^{\int_{a+2\pi i n/\beta}^{a+2\pi i (n+1)/\beta} dx (-S_{\mathrm{odd}}+f)}
    \right]
    (1 - A)
    \notag\\&\qquad\times
    \left[\prod_{n=0}^{\infty}
    e^{\int_{-a-2\pi i (n+1)/\beta}^{-a-2\pi i n/\beta} dx (-S_{\mathrm{odd}}+f)}
    \right]
    e^{\int_{-a-\infty}^{-a-i\infty} dx (-S_{\mathrm{odd}}+f)}
    \label{eq:zel_quantization_0}
    \\
    &= e^{\int_{a}^{-a} dx (-S_{\mathrm{odd}}+f)} (1-A),
    \label{eq:zel_quantization}
\end{align}
where the turning point $a=\beta^{-1}\cosh^{-1}\sqrt{U_0/E}$.
In the second equality, we have used some properties shown in Fig.~\ref{fig:zel_integration}.
Note that if $\varepsilon=0$ (the usual resonant state) then the integration over $x\in(-\infty,\infty)$ cannot be finite because $e^{\int_{\pm\infty}^{\pm\infty\pm i\infty}dx\, S_{\mathrm{odd}}}$ diverges, so we should take the other contour given in~Ref.~\cite{Morikawa:2025grx}.\footnote{When $\varepsilon>0$, we can take an \textit{asymmetric} path at infinity, but one finds an overall factor that is irrelevant to the quantization condition. Equation~\eqref{eq:zel_quantization} implies that the path is anti-symmetric.}

\begin{figure}
    \centering
\begin{tikzpicture}[scale=0.9]
    \draw[->] (-5,0) -- (5,0) node[right] {$\re x$};
    \draw[->] (0,-5) -- (0,5) node[above] {$\im x$};
    \draw[very thick,->,red] (-5.5,-0.5) -- (-5.5,-5.5) -- (-2,-5.5) -- (-2,0) -- (2,0) -- (2,5.5) -- (5.5,5.5) -- (5.5,0.5);
    \fill (-2,0) circle(5pt) node[above right] {$-a$}; \fill (2,0) circle(5pt) node[above right] {$a$};
    \fill (-2,-2) circle(5pt) node[above left] {$(-a,2\pi n)$}; \fill (2,2) circle(5pt) node[below right] {$(a,2\pi n)$};
    \fill (-2,-3) circle(5pt) node[below left] {$(-a,2\pi(n+1))$}; \fill (2,3) circle(5pt) node[above right] {$(a,2\pi(n+1))$};
    \draw[ultra thick,->,blue] (-2,-3) -- (-2,-2) node[midway,left] {$e^{I_n}=e^{\int S_{\mathrm{odd}}+f}$};
    \draw[ultra thick,->,blue] (2,2) -- (2,3) node[midway,right] {$e^{-I_n}$};
    \draw[ultra thick,->,orange] (-5.5,-5.6) -- (-2,-5.6) node[right] {$e^{I_{i\infty}}=e^{\int S_{\mathrm{odd}}}$};
    \draw[ultra thick,->,orange] (2,5.6) -- (5.5,5.6) node[right] {$e^{-I_{i\infty}}$};
    \draw[ultra thick,->,red!40] (-5.6,-0.5) -- (-5.6,-5.5) node[midway,left] {$e^{I_\infty}=e^{\int S_{\mathrm{odd}}+f}\to0$};
    \draw[ultra thick,->,red!40] (5.6,5.5) -- (5.6,0.5) node[midway,right] {$e^{-I_\infty}$};
\end{tikzpicture}
    \caption{Illustration of computing Eq.~\eqref{eq:zel_quantization}. $\beta=1$. Although $f(x)$ diverges near $|\im x|\to\infty$, $f(-x)=-f(x)$ and hence $f(x)$ does not contribute to~$e^{I_{i\infty}}$.
    Note that $I_n$, $I_{i\infty}$ and $I_\infty$ are finite.
    (If $f=0$ then $e^{I_\infty}$ is divergent.)}
    \label{fig:zel_integration}
\end{figure}

The quantum system with the (inverted) Rosen--Morse potential is exactly solvable.
The solution is given by
\begin{align}
    \psi(x) = e^{-\varepsilon x^2}
    (1-\xi^2)^{-\frac{ik}{2\beta}}
    F\left( -\frac{ik}{\beta}-s, -\frac{ik}{\beta}+s+1, -\frac{ik}{\beta}+1, \frac{1-\xi}{2} \right) ,
 \label{eq:rm_sol}
\end{align}
where the first factor depending on~$\varepsilon$ is the Zel'dovich regulator;
$\xi = \tanh \beta x$
\begin{align}
    k = \frac{\sqrt{2mE}}{\hbar}, \quad
    -\frac{2mU_{0}}{\beta^2\hbar^2} = s(s+1), \quad
    s=\frac{1}{2}\left( -1+\sqrt{1-\frac{8mU_{0}}{\beta^2 \hbar^2}} \right) ,
\end{align}
and $F$ is the Gauss hypergeometric function.
Therefore, we can compute the exact expression in~Eq.~\eqref{eq:zel_quantization} as (we can take the limit~$\varepsilon\to0$); the formula of $A$-cycle is explicitly given in Ref.~\cite{Morikawa:2025grx} and one finds
\begin{align}
    \left[
    \frac{F\left( -\frac{ik}{\beta}-s, -\frac{ik}{\beta}+s+1, -\frac{ik}{\beta}+1, \frac{1-|\xi|}{2} \right)}
    {F\left( -\frac{ik}{\beta}-s, -\frac{ik}{\beta}+s+1, -\frac{ik}{\beta}+1, \frac{1+|\xi|}{2} \right)}
    \right]^2
    = 1 ,
    \label{eq:quant_cond_F}
\end{align}
where $|\xi|=\tanh(\cosh^{-1}\sqrt{U_0/E})$.
Noting the formula of the hypergeometric function as
\begin{align}
    &F\left( -\frac{ik}{\beta}-s, -\frac{ik}{\beta}+s+1, -\frac{ik}{\beta}+1, \frac{1\pm|\xi|}{2} \right)\notag\\
    &= \mathfrak{A} F\left(-\frac{ik}{2\beta}-\frac{s}{2},-\frac{ik}{2\beta}+\frac{s+1}{2},\frac{1}{2},|\xi|^2\right)
    \mp \mathfrak{B} F\left(-\frac{ik}{2\beta}-\frac{s-1}{2},-\frac{ik}{2\beta}+\frac{s}{2}+1,\frac{3}{2},|\xi|^2\right) ,
\end{align}
where
\begin{align}
    \mathfrak{A} &= \frac{\Gamma\left(-\frac{ik}{\beta}+1\right)\Gamma\left(\frac{1}{2}\right)}
    {\Gamma\left(-\frac{ik}{2\beta}-\frac{s-1}{2}\right)\Gamma\left(-\frac{ik}{2\beta}+\frac{s}{2}+1\right)}, &
    \mathfrak{B} &= \frac{\Gamma\left(-\frac{ik}{\beta}+1\right)\Gamma\left(-\frac{1}{2}\right)}
    {\Gamma\left(-\frac{ik}{2\beta}-\frac{s}{2}\right)\Gamma\left(-\frac{ik}{2\beta}+\frac{s+1}{2}\right)} ,
\end{align}
we observe the complex energy for resonance with $\frac{8mU_{0}}{\beta^2 \hbar^2}>1$
\begin{align}
    E=\frac{\hbar^2 \beta^2}{8m}
    \left[\sqrt{\frac{8mU_{0}}{\beta^2 \hbar^2}-1}-i(2n+1)\right]^2,
    \qquad n\in\mathbb{Z}_{\geq0}.
    \label{eq:res-ene}
\end{align}

\subsection{Remarks on difficulties of irrational potential}
We make remarks on some technical and conceptual points.
Figure~\ref{fig:stokes_graph} corresponds to $\im\hbar\gtrless0$ and $\im E\gtrless0$, while Fig.~\ref{fig:stokes_graph2} is $\im\hbar\gtrless0$ and $\im E\lessgtr0$.
In the case that $\im\hbar>0$, originally, for $\im E>0$, $\psi^{+}$ beginning at the left turning point, $-a$, is absorbed into the above double pole. However, for $\im E<0$, the curve of $\psi^{+}$ becomes flat and so falls down.

Actually, this phenomenon is quite difficult since we focus on non-polynomial and irrational potential functions such as the inverted Rosen--Morse potential. It is not straightforward to consider the monodromy and connection formula around the double poles.
In fact, for an irrational potential, the residue theorem can be violated; it makes the derivation of explicit connection formula hard since estimations of some contour integrals become too nontrivial. To our best knowledge, in particular, the Stokes phenomenon across the Stokes curve connecting the two turning points in Fig.~\ref{fig:stokes_graph2} is subtle. Furthermore, though we mentioned $\im E\sim0$, for large $|\im E|$ there are two Stokes curves extending from a double pole to $\im x\to\pm\infty$. The monodromy associated with these curves may be unidentifiable in an analytic sense due to such an irrational potential.\footnote{The validity of our analytical strategy should be examined through numerical verification in future studies.}

\section{Complex scaling method}
\subsection{Solving method within complex scaling method}
For the inverted Rosen–Morse potential, the Schr\"{o}dinger equation with the complex scaling method (CSM) is given by
\begin{align}
    \left[-\frac{\hbar^2}{2m} \frac{d^2}{dx'^2} + \frac{U_{0}}{\cosh^2 \beta x'}\right] \psi(x')
    = E \psi(x') ,
\end{align}
where we have replaced $x\in\mathbb{R}$ by $x'=xe^{i\theta}\in\mathbb{C}$ with the scaling angle $\theta$. The range of $\theta$ is restricted to $0<\theta<\pi/4$ in the CSM. Using $\xi=\tanh\beta x'$, the equation is represented as
\begin{align}
    \left[ \frac{d}{d\xi} (1-\xi^2)\frac{d}{d\xi} +s(s+1) - \frac{\kappa^2}{1-\xi^2} \right] \psi(x')
    &= 0 , \qquad
    \kappa = \frac{\sqrt{-2mE}}{\beta\hbar} .
    \label{eq:CSM-reduced-eq}
\end{align}
Letting $\psi(x')=(1-\xi^2)^{\frac{\kappa}{2}}\omega(\xi)$ and substituting this into Eq.~\eqref{eq:CSM-reduced-eq}, we obtain the following equation:
\begin{align}
    \left[
    u(1-u)\frac{d^2}{d\xi^2} + (\kappa + 1) (1 - 2 u) \frac{d}{d\xi} - (\kappa - s) (\kappa + s + 1) \right] \omega(\xi) = 0, \quad u=\frac{1-\xi}{2} .
\end{align}
The solution of this equation, which is regular at $x=0$, is given by
\begin{align}
    \psi_{\mathrm{CSM}}(x)= (1-\xi^2)^{\frac{\kappa}{2}} 
    F\left( \kappa-s, \kappa+s+1, \kappa+1, \frac{1-\xi}{2} \right).
\end{align}
In general, as $x\to-\infty$ ($\xi\to-1$), the Gauss hypergeometric function diverges. However, physical wave functions of resonant states under CSM are normalizable because of the ABC theorem~\cite{Aguilar:1971ve,Balslev:1971vb}. To ensure the normalizability of the wave function, the hypergeometric function should be expressed as a finite-order polynomial. This requirement leads to the condition, $\kappa-s=-n$ $(n\in\mathbb{Z}_{\geq0})$, which gives
\begin{align}
    E = \frac{\hbar^2\beta^2}{8m} \left[\sqrt{\frac{8mU_0}{\beta^2 \hbar^2} - 1} - i(2n + 1) \right]^2.
\end{align}
This complex energy is completely the same result as Eq.~\eqref{eq:res-ene}.

\subsection{Exact WKB framework of complex scaling method}
The potential~$Q$ of the CSM Schr\"odinger equation can be rewritten by
\begin{align}
    Q_{\mathrm{CSM}}(x) = 2e^{2i\theta}\left(\frac{U_{0}}{\cosh^2 \beta xe^{i\theta}} - E\right) .
\end{align}
The Stokes graph, i.e., Fig.~\ref{fig:stokes_graph}, is rotated by~$\theta$.
Now, the path for the quantization condition can be schematically depicted as in~Fig.~\ref{fig:csm_integration}.
Thus, we have the almost same quantization condition as~Eq.~\eqref{eq:zel_quantization_0} with~$f(x)=0$, but the interval of the integration $\int S_{\mathrm{odd}}$ is smaller as follows:
\begin{align}
    \lim_{r\to\infty}\int_{\pm(1+i)r}^{\pm r} S_{\mathrm{odd}}
    \to 
    \lim_{r\to\infty}\int_{\pm(1+i)r}^{\pm e^{\pm i\theta}r} S_{\mathrm{odd}}^\theta .
\end{align}
The CSM excludes the region near~$x=\pm\infty$, where the solution is most singular.
At $\re x\to\infty$, the analytic continuation is supposed to be carried out from the argument $\pi/4$ to $\theta$.
Therefore, from the formula in Ref.~\cite{Morikawa:2025grx}, we have the overall convergent factor in the quantization condition as
\begin{align}
    &\lim_{r\to\infty}\exp{\int_{e^{i\theta}r}^{e^{i\pi/4}r} S_{\mathrm{odd}}^\theta}\notag\\
    &=\lim_{r\to\infty}
    \frac{F\left( \kappa-s, \kappa+s+1, \kappa+1, \frac{1-\tanh\beta r e^{i\pi/4}}{2} \right)}
    {F\left( \kappa-s, \kappa+s+1, \kappa+1, \frac{1-\tanh\beta r e^{i\theta}}{2} \right)}
    = \text{finite}
    \qquad \text{if $0<\theta<\frac{\pi}{4}$} .
\end{align}
The non-perturbative cycle $A$ is now given by
\begin{align}
    A = \exp\oint_{-a e^{-i\theta}}^{a e^{-i\theta}} S_{\mathrm{odd}}^\theta
    = \left[\frac{\psi_{\mathrm{CSM}}(a e^{-i\theta})}{\psi_{\mathrm{CSM}}(-a e^{-i\theta})}\right]^2
    = \left[\frac{\psi(a)}{\psi(-a)}\right]^2
    = \exp\oint_{-a}^a S_{\mathrm{odd}} ,
\end{align}
where $\psi_{\mathrm{CSM}}(x' e^{-i\theta}) = \psi(x')$, $a e^{-i\theta}$ denotes the turning point after complex rotation, and $a$ is the original turning point with $\theta=0$.
We obtain the well-defined complex energy being identical to~Eq.~\eqref{eq:res-ene}.

\begin{figure}[t]
    \centering
\begin{tikzpicture}[scale=0.9]
    \draw[->] (-7,0) -- (7,0) node[right] {$\re x$};
    \draw[->] (0,-4) -- (0,4) node[above] {$\im x$};
    \draw[very thick,->,orange,rotate around={-22.5:(0,0)}] (-5.5,0) -- (-5.5,-5.5) -- (-2,-5.5) -- (-2,0) -- (2,0) -- (2,5.5) -- (5.5,5.5) -- (5.5,0);
    \fill[rotate around={-22.5:(0,0)}] (-2,0) circle(5pt) node[above right] {$-a e^{-i\theta}$}; \fill[rotate around={-22.5:(0,0)}] (2,0) circle(5pt) node[below right] {$a e^{-i\theta}$};
    \fill[blue,rotate around={-22.5:(0,0)}] (0,1) circle(3pt); \fill[blue,rotate around={-22.5:(0,0)}] (0,3) circle(3pt);
    \fill[blue,rotate around={-22.5:(0,0)}] (0,-1) circle(3pt); \fill[blue,rotate around={-22.5:(0,0)}] (0,-3) circle(3pt);
    \draw[dotted,blue,rotate around={-22.5:(0,0)}] (0,-4) -- (0,4);
    \draw[blue] (0,1) arc(90:67.5:1) node[midway,above] {$\theta$};
    \draw[ultra thick,->,green!50!black,rotate around={-22.5:(0,0)}] (-5.6,-2.3) -- (-5.6,-5.5) node[midway,left] {$e^{I'_\infty}=e^{\int S_{\mathrm{odd}}}$}; \fill[green!50!black,rotate around={-22.5:(0,0)}] (-5.6,-2.3) circle(5pt) node[above left] {$-\infty$};
    \draw[ultra thick,->,green!50!black,rotate around={-22.5:(0,0)}] (5.6,5.5) -- (5.6,2.4) node[midway,right] {$e^{-I'_\infty}$}; \fill[green!50!black,rotate around={-22.5:(0,0)}] (5.6,2.3) circle(5pt) node[below right] {$\infty$};
\end{tikzpicture}
    \caption{Rotated path on the inverted Rosen--Morse potential by~$\theta$.
    The black points and the blue points denote the turning points and the double poles, respectively.
    The orange path is the usual contour ($\theta=0$) given in~Figs.~\ref{fig:stokes_graph_path} and \ref{fig:zel_integration},
    which is rotated by~$\theta$ under the CSM.
    The endpoints of the path in the CSM are at~$x=\pm\infty$, and so the interval of the integration $\int S_{\mathrm{odd}}$ is smaller than the usual one by~$\theta$.
    $I'_\infty$ is finite for $0<\theta<\pi/4$.}
    \label{fig:csm_integration}
\end{figure}

\section{Rigged Hilbert space}
Finally, we define a Hilbert space on the exact WKB framework.
In the usual context of the exact WKB analysis, details of the Hilbert space of WKB solutions are not mentioned explicitly.
In Table~\ref{tab:ewkb_space}, we summarize the classification of the corresponding linear spaces defined in each step of the exact WKB manipulations. This can be explained as follows.

\begin{table}[t]
    \centering
\begin{tabular}{ll}\toprule
    Wave function & Linear space\\\midrule
    Formal power series & Formal infinite-dim linear space\\
        &\, (not convergent)\\
    Borel resummation & (Infinite-dim) Function space\\
        &\, (convergent/analytically-continuable\\
        &\quad in each Borel summable region) \\
    Quantization & Norm space\\
        &\, \textcolor{red}{but Banach space (completeness) if bound state}\\
    Observable & Metric space (inner product)\\
        &\, \textcolor{red}{but Hilbert space (completeness) if bound state}\\
    \midrule
    Resonance & At most function space\\
    Resonance regularized & ``Hilbert space'' where resonance is bound state\\
        &\, \textcolor{red}{with different def of norm/inner product}\\
        &\, (completeness)\\
    Regularization off & Unbound state\\
    \bottomrule
\end{tabular}
    \caption{Classification of linear space in each step of exact WKB analysis.}
    \label{tab:ewkb_space}
\end{table}

At first, the space of the WKB ansatz, i.e., formal power series~$\psi(x,\hbar)^{\pm}$, is just a formal linear space, where we have only the formal observables as position ($X\in\mathbb{C}$), momentum ($P$), and the Hamiltonian ($H$), and no other mathematical manipulations are defined.
After the Borel resummation, the Borel integral~$\Psi(x,\hbar)^{\pm}$ is convergent and analytically continuable if $\Psi(x,\hbar)^{\pm}\in\mathcal{D}_{\mathcal{B}}[Q(x),\hbar]$ where $\mathcal{D}_{\mathcal{B}}$ is a Borel-summable region connected to~$x$ (recall that~Eq.~\eqref{eq:stokes_curve}).
Thus, in~$\mathcal{D}_{\mathcal{B}}|_{x}$, the set of wave functions build up a function space.
Also, we can use the monodromy matrix~$M_{\pm}$ across a Stokes curve and the connection formula between turning points, and hence $\cup\mathcal{D}_{\mathcal{B}}$ is well-defined on $X\in\mathbb{C}$.

Now, we assume that some kind of physical condition holds.
For instance, we have introduced the quantization condition;
we have defined the norm of wave functions which should be convergent in a physical sense.
This means that our linear space is supposed to be a norm space.
If there exist only bound states, it is straightforward to prove the completeness and then this is a Banach space.
Then, for computing any observable, we have the metric space or Hilbert space if bound state.

Resonance (or scattering) is remarkable because such a quasi-stable state is not normalizable; we should use regularization such as the CSM.
Also, the probability becomes a complex number~\cite{Berggren:1996}, and a transition cross-section is quite subtle~\cite{Berggren:1971,Berggren:1978}.
We have already shown the equivalence and complementarity between the exact WKB analysis, Zel'dovich regularization, and the CSM.
The most important point is that the crucial singular region
\begin{align}
    \mathcal{D}_\varepsilon^{\mathrm{R}}
    = \{ x\in\mathbb{C} | \varepsilon>0, \lim_{r\to\pm\infty}|x - r| < \varepsilon \}
\end{align}
should be eliminated, and so $X\in\mathbb{C}\setminus\mathcal{D}_\varepsilon^{\mathrm{R}}$.
Then, the ABC theorem of the CSM (the completeness of the Hilbert space) is valid in the exact WKB framework for resonance if $X\in\mathbb{C}\setminus\mathcal{D}_\varepsilon^{\mathrm{R}}$.
Therefore, we have the well-defined Hilbert space including resonance, $\mathcal{H}_\varepsilon$ with~$\varepsilon>0$.
Note that the regulator~$\varepsilon$ indicates that the concrete definition of the norm or inner product is quite different as shown in the above figures.

To observe some physical quantities, we consider a set of operators, $\{A_i\}$.
$A_i$ must be defined on the domain $D(A_i)\subset\mathcal{H}_\varepsilon$; then let us introduce the dense subspace
\begin{align}
    \Phi \equiv \cap_i D(A_i) \subseteq \mathcal{H}_\varepsilon .
\end{align}
In our prescription, truly unbound states as resonance can be realized by the range/codomain with the limit $\varepsilon\to0$, $\Phi^\times$.
Due to the elimination of the regulator, $\mathcal{H}_\varepsilon\subset\Phi^\times$.
Therefore we have the Gelfand triplet~\cite{DelaMadrid:2001dln,DelaMadrid:2001wwa,DelaMadrid:2002cz}
\begin{align}
    \Phi\subseteq\mathcal{H}_\varepsilon\subset\Phi^\times ,
\end{align}
and this modified Hilbert space with the pair~$(\mathcal{H}_\varepsilon,\Phi)$ is called the rigged Hilbert space.

\section{Conclusion}
We have proposed a unified exact WKB framework to describe quantum mechanical resonances, incorporating both the Zel'dovich regularization and complex scaling method. Through the case study of the inverted Rosen--Morse potential, we confirmed that exact WKB analysis not only reproduces known resonance structures but also provides a coherent picture of their analytic properties.

Our analysis demonstrates that different regularization schemes are consistent with one another when interpreted through the lens of exact WKB theory.
Additionally, we also found the construction of the rigged Hilbert space in the exact WKB method; we saw the most essential regulator inhabiting every regularization of resonant states.
This finding strengthens the validity of exact WKB analysis as a robust method for studying quasi-stationary states beyond bound systems.

Looking forward, our framework paves the way for broader applications of exact WKB theory in non-Hermitian quantum mechanics, open quantum systems, and possibly quantum field theory. It may also serve as a bridge between mathematical rigor and physical intuition in the study of unstable phenomena in quantum systems.

\section*{Acknowledgements}
We are grateful to the referees for helpful advice and a careful reading
of the manuscript.
We would like to thank the officers in Nishijin Plaza at Kyushu University, and \mbox{Izakaya En} for their hospitality.
This work was partially supported by Japan Society for the Promotion of Science (JSPS)
Grant-in-Aid for Scientific Research Grant Numbers
JP22KJ2096, JP25K17402 (O.M.) and JP21H04975 (S.O.).
O.M.\ acknowledges the RIKEN Special Postdoctoral Researcher Program
and RIKEN FY2025 Incentive Research Projects.

\bibliographystyle{utphys}
\bibliography{ref,ref_ewkb,ref_res}

\providecommand{\href}[2]{#2}\begingroup\raggedright\begin{thebibliography}{10}

\bibitem{Voros:1983xx}
A.~Voros, ``The return of the quartic oscillator. the complex wkb method,'' {\em Ann. Inst. Henri Poincar\'e} {\bfseries 39} no.~3, (1983) 211–338.

\bibitem{Delabaere:1999xx}
E.~Delabaere and F.~Pham, ``Resurgent methods in semi-classicaasymptotics,'' {\em Ann. Inst. Henri Poincar\'e} {\bfseries 71} (1999) 1--94.

\bibitem{Iwaki:2014vad}
K.~Iwaki and T.~Nakanishi, ``{Exact WKB analysis and cluster algebras},'' \href{http://dx.doi.org/10.1088/1751-8113/47/47/474009}{{\em J. Phys. A} {\bfseries 47} no.~47, (2014) 474009}, \href{http://arxiv.org/abs/1401.7094}{{\ttfamily arXiv:1401.7094 [math.CA]}}.

\bibitem{Morikawa:2025grx}
O.~Morikawa and S.~Ogawa, ``{Non-perturbative formulation of resonances in quantum mechanics based on exact WKB method},'' \href{http://arxiv.org/abs/2503.18741}{{\ttfamily arXiv:2503.18741 [hep-th]}}.

\bibitem{Sueishi:2020rug}
N.~Sueishi, S.~Kamata, T.~Misumi, and M.~\"Unsal, ``{On exact-WKB analysis, resurgent structure, and quantization conditions},'' \href{http://dx.doi.org/10.1007/JHEP12(2020)114}{{\em JHEP} {\bfseries 12} (2020) 114}, \href{http://arxiv.org/abs/2008.00379}{{\ttfamily arXiv:2008.00379 [hep-th]}}.

\bibitem{Sueishi:2021xti}
N.~Sueishi, S.~Kamata, T.~Misumi, and M.~\"Unsal, ``{Exact-WKB, complete resurgent structure, and mixed anomaly in quantum mechanics on S$^{1}$},'' \href{http://dx.doi.org/10.1007/JHEP07(2021)096}{{\em JHEP} {\bfseries 07} (2021) 096}, \href{http://arxiv.org/abs/2103.06586}{{\ttfamily arXiv:2103.06586 [quant-ph]}}.

\bibitem{Kamata:2021jrs}
S.~Kamata, T.~Misumi, N.~Sueishi, and M.~\"Unsal, ``{Exact WKB analysis for SUSY and quantum deformed potentials: Quantum mechanics with Grassmann fields and Wess-Zumino terms},'' \href{http://dx.doi.org/10.1103/PhysRevD.107.045019}{{\em Phys. Rev. D} {\bfseries 107} no.~4, (2023) 045019}, \href{http://arxiv.org/abs/2111.05922}{{\ttfamily arXiv:2111.05922 [hep-th]}}.

\bibitem{Zhao:1995}
M.~Zhao, ``Dynamical resonances and lifetimes in rigged hilbert spaces,'' \href{http://dx.doi.org/https://doi.org/10.1016/0375-9601(95)00510-A}{{\em Phys. Lett. A} {\bfseries 204} (1995) 319--322}.

\bibitem{Landau:1991wop}
L.~D. Landau and E.~M. Lifshits, {\em {Quantum Mechanics}: {Non-Relativistic Theory}}, vol.~v.3 of {\em Course of Theoretical Physics}.
\newblock Butterworth-Heinemann, Oxford, 1991.

\bibitem{Ryaboy:1993}
V.~Ryaboy and N.~Moiseyev, ``Cumulative reaction probability from siegert eigenvalues: Model studies,'' \href{http://dx.doi.org/10.1063/1.464392}{{\em The Journal of Chemical Physics} {\bfseries 98} no.~12, (06, 1993) 9618--9623}, \href{http://arxiv.org/abs/https://pubs.aip.org/aip/jcp/article-pdf/98/12/9618/19327646/9618\_1\_online.pdf}{{\ttfamily https://pubs.aip.org/aip/jcp/article-pdf/98/12/9618/19327646/9618\_1\_online.pdf}}.

\bibitem{Zel'dovich:1961}
Y.~B. Zel'dovich, ``{On the theory of unstable states},'' {\em Zh. \'Eksp. Teor. Fiz.} {\bfseries 39} (1961) 776. [Sov. Phys. JETP 12, 542 (1961)].

\bibitem{Berggren:1968zz}
T.~Berggren, ``{On the use of resonant states in eigenfunction expansions of scattering and reaction amplitudes},'' \href{http://dx.doi.org/10.1016/0375-9474(68)90593-9}{{\em Nucl. Phys. A} {\bfseries 109} (1968) 265--287}.

\bibitem{Aguilar:1971ve}
J.~Aguilar and J.~M. Combes, ``{A class of analytic perturbations for one-body schroedinger hamiltonians},'' \href{http://dx.doi.org/10.1007/BF01877510}{{\em Commun. Math. Phys.} {\bfseries 22} (1971) 269--279}.

\bibitem{Balslev:1971vb}
E.~Balslev and J.~M. Combes, ``{Spectral properties of many-body schroedinger operators with dilatation-analytic interactions},'' \href{http://dx.doi.org/10.1007/BF01877511}{{\em Commun. Math. Phys.} {\bfseries 22} (1971) 280--294}.

\bibitem{Myo:2014ypa}
T.~Myo, Y.~Kikuchi, H.~Masui, and K.~Kat\={o}, ``{Recent development of complex scaling method for many-body resonances and continua in light nuclei},'' \href{http://dx.doi.org/10.1016/j.ppnp.2014.08.001}{{\em Prog. Part. Nucl. Phys.} {\bfseries 79} (2014) 1--56}, \href{http://arxiv.org/abs/1410.4356}{{\ttfamily arXiv:1410.4356 [nucl-th]}}.

\bibitem{Rafael:2012}
R.~de~la Madrid, ``{The rigged Hilbert space approach to the Gamow states},'' \href{http://dx.doi.org/10.1063/1.4758925}{{\em J. Math. Phys.} {\bfseries 53} (2012) 102113}.

\bibitem{Antoniou:2001}
I.~Antoniou, Y.~Melnikov, and E.~Yarevsky, ``The connection between the rigged hilbert space and the complex scaling approaches for resonances. the friedrichs model,'' \href{http://dx.doi.org/https://doi.org/10.1016/S0960-0779(01)00082-0}{{\em Chaos, Solitons \& Fractals} {\bfseries 12} (2001) 2683--2688}.

\bibitem{Kamata:2023opn}
S.~Kamata, ``{Exact WKB analysis for PT-symmetric quantum mechanics: Study of the Ai-Bender-Sarkar conjecture},'' \href{http://dx.doi.org/10.1103/PhysRevD.109.085023}{{\em Phys. Rev. D} {\bfseries 109} no.~8, (2024) 085023}, \href{http://arxiv.org/abs/2401.00574}{{\ttfamily arXiv:2401.00574 [hep-th]}}.

\bibitem{Kamata:2024tyb}
S.~Kamata, ``{Exact quantization conditions and full transseries structures for PT symmetric anharmonic oscillators},'' \href{http://dx.doi.org/10.1103/PhysRevD.110.045022}{{\em Phys. Rev. D} {\bfseries 110} no.~4, (2024) 045022}, \href{http://arxiv.org/abs/2406.01230}{{\ttfamily arXiv:2406.01230 [hep-th]}}.

\bibitem{Nikolaev:2008}
N.~Nikolaev, ``{Exact Solutions for the Singularly Perturbed Riccati Equation and Exact WKB Analysis},'' \href{http://dx.doi.org/10.1017/nmj.2022.38}{{\em Nagoya Mathematical Journal} {\bfseries 250} (2023) 434--469}, \href{http://arxiv.org/abs/2008.06492}{{\ttfamily 2008.06492 [math.CA]}}.

\bibitem{Nikolaev:2021xzt}
N.~Nikolaev, ``{Existence and Uniqueness of Exact WKB Solutions for Second-Order Singularly Perturbed Linear ODEs},'' \href{http://dx.doi.org/10.1007/s00220-022-04603-7}{{\em Commun. Math. Phys.} {\bfseries 400} no.~1, (2023) 463--517}, \href{http://arxiv.org/abs/2106.10248}{{\ttfamily arXiv:2106.10248 [math.AP]}}.

\bibitem{Moiseyev:2011}
N.~Moiseyev, {\em {Non-Hermitian Quantum Mechanics}}.
\newblock Cambridge University Press, Cambridge, 2011.

\bibitem{Berggren:1996}
T.~Berggren, ``Expectation value of an operator in a resonant state,'' \href{http://dx.doi.org/https://doi.org/10.1016/0370-2693(96)00132-3}{{\em Physics Letters B} {\bfseries 373} no.~1, (1996) 1--4}.

\bibitem{Berggren:1971}
T.~Berggren, ``On the treatment of resonant final states in direct reactions,'' \href{http://dx.doi.org/https://doi.org/10.1016/0375-9474(71)90889-X}{{\em Nuclear Physics A} {\bfseries 169} no.~2, (1971) 353--362}.

\bibitem{Berggren:1978}
T.~Berggren, ``On the interpretation of complex cross sections for production of resonant final states,'' \href{http://dx.doi.org/https://doi.org/10.1016/0370-2693(78)90747-5}{{\em Physics Letters B} {\bfseries 73} no.~4, (1978) 389--392}.

\bibitem{DelaMadrid:2001dln}
R.~de~la Madrid, ``{Rigged Hilbert Space Approach to the Schrodinger Equation},'' \href{http://dx.doi.org/10.1088/0305-4470/35/2/311}{{\em J. Phys. A} {\bfseries 35} (2002) 319--342}, \href{http://arxiv.org/abs/quant-ph/0110165}{{\ttfamily arXiv:quant-ph/0110165}}.

\bibitem{DelaMadrid:2001wwa}
R.~de~la Madrid, A.~Bohm, and M.~Gadella, ``{Rigged Hilbert Space Treatment of Continuous Spectrum},'' \href{http://dx.doi.org/10.1002/1521-3978(200203)50:2<185::AID-PROP185>3.0.CO;2-S}{{\em Fortsch. Phys.} {\bfseries 50} (2002) 185--216}, \href{http://arxiv.org/abs/quant-ph/0109154}{{\ttfamily arXiv:quant-ph/0109154}}.

\bibitem{DelaMadrid:2002cz}
R.~de~la Madrid and M.~Gadella, ``{A Pedestrian introduction to Gamow vectors},'' \href{http://dx.doi.org/10.1119/1.1466817}{{\em Am. J. Phys.} {\bfseries 70} (2002) 626--638}, \href{http://arxiv.org/abs/quant-ph/0201091}{{\ttfamily arXiv:quant-ph/0201091}}.

\end{thebibliography}\endgroup
\end{document}